\documentstyle[11pt,newpasp,twoside,epsf]{article}
\def\edcomment#1{\iffalse\marginpar{\raggedright\sl#1\/}\else\relax\fi}
\reversemarginpar

\begin{document}
\title{Highly-Ionized Intergalactic Gas at Low Redshifts: Constraints from 
QSO Absorption Lines}
\author{Todd M. Tripp}
\affil{Princeton University Observatory, Peyton Hall, Princeton, NJ 08544}


\begin{abstract}
High-resolution UV spectroscopy of low$-z$ QSOs with the Space Telescope Imaging 
Spectrograph and the {\it Far Ultraviolet Spectroscopic Explorer} has indicated 
that O~VI absorption-line systems provide a valuable probe of the low$-z$ 
intergalactic medium. These observations and their implications are briefly 
reviewed. Though still uncertain due to the small sample, the number of O~VI 
absorbers per unit redshift is quite high, and these absorbers appear to trace 
an important baryon reservoir. The O~VI systems are {\it intervening}; they are 
highly displaced from the background QSO redshifts and are correlated with 
foreground luminous galaxies. Their physical conditions are variable and 
sometimes complicated. In some cases, there is clear evidence that the absorbers 
have a multiphase nature. Some appear to be photoionized by the UV background 
from QSOs, while others are probably collisionally ionized. However, the 
majority of the O~VI lines have $b-$values that suggest an origin in 
collisionally ionized hot gas. The observations are in agreement with 
hydrodynamic simulations of cosmological structure growth, which predict 
substantial quantities of shock-heated, hot intergalactic gas at low $z$, but 
more observational and theoretical work is needed to confirm the nature of the 
O~VI systems.
\end{abstract}

\section{Introduction}

Understanding the quantity, distribution, and physical state of the baryons 
in the universe is an important problem in current cosmology.  Hydrodynamic 
simulations of cosmological structure growth predict that in the nearby 
universe, 30-50\% of the baryons are in shock-heated intergalactic gas at 
$10^{5} - 10^{7}$ K (Cen \& Ostriker 1999; Dav\'{e} et al. 2001). Gas with $T < 
10^{6}$ K may be readily detected using O~VI absorption lines in the spectrum of 
a background QSO.  Observations with the {\it Hubble Space Telescope} Imaging 
Spectrograph (STIS) and the {\it Far Ultraviolet Spectroscopic Explorer (FUSE)} 
have recently revealed a surprising number of low-redshift O~VI absorption line 
systems in the directions of a few QSOs, and the inferred high number of O~VI 
systems per unit redshift suggests that these absorbers are indeed an 
important baryon reservoir.  These observations and their implications 
regarding the physical nature and distribution of the intergalactic gas as 
well as its baryonic content are reviewed. This brief review attempts to collate 
results from collaborations with several groups involving major contributions 
from many people. Table 1 summarizes the publications from the first batch of 
QSOs that have been observed and provides some basic information on the 
intervening O~VI lines detected in each sight line. After a few notes on the 
observations and data involved (\S 2), some comments are provided on the O~VI 
absorber properties (\S 3), the baryonic content of these absorption systems (\S 
4), and the ionization and physical conditions of the gas (\S 5). The paper 
is summarized in \S 6.

\section{Observations}

\begin{table}[h]
\vspace*{-0.3cm}
\small
\begin{center}
\begin{tabular}{lcclll}
\multicolumn{6}{c}{Table 1: Summary of First Detections of Low-z Intervening 
O~VI Absorbers} \\
\multicolumn{6}{c}{with STIS and FUSE} \\ \hline \hline
Sight Line & $z_{\rm QSO}$ & $z_{\rm abs}$(O VI)$^{a}$ & $W_{\rm r}^{b}$ & 
Class$^{c}$ & Reference$^{d}$ \\
  \ & \ & \ & (m\AA ) & \ & \ \\ \hline
PKS0405-123 & 0.573 & 0.16710 & 491$\pm$46 & Single-component & 1,2 \\
   \        &   \   & 0.18290 & 94$\pm$23  & Single-component & 1 \\
   \        &   \   & 0.36332 & 41$\pm$7   & (Single-component) & 1 \\
   \        &   \   & 0.49512 & 225$\pm$15 & Multicomponent   & 1 \\
PG0953+415  & 0.239 & 0.06807 & 127$\pm$9  & Single-component & 3 \\
   \        &   \   & 0.14232 & 112$\pm$12 & Single-component & 3,4 \\
3C 273      & 0.158 & 0.12004 & 28$\pm$5   & Single-component & 5,1 \\
3C 351      & 0.372 & 0.22110 & ...$^{e}$  & (Single-component) & 1 \\
   \        &   \   & 0.31656 & 230$\pm$14 & Multicomponent & 1 \\
H1821+643   & 0.297 & 0.12120 & 90$\pm$17  & Single-component & 6,7 \\
   \        &   \   & 0.21326 & 38$\pm$9   & (Single-component) & 8 \\
   \        &   \   & 0.22497 & 185$\pm$9  & Multicomponent   & 8 \\
   \        &   \   & 0.22637 & 25$\pm$5   & Multicomponent$^{f}$   & 8 \\
   \        &   \   & 0.24531 & 55$\pm$6   & Single-component & 8 \\
   \        &   \   & 0.26659 & 55$\pm$8   & Single-component & 8 \\
\hline
\end{tabular}
\end{center}
$^{a}$Heliocentric absorption system redshift. \\
$^{b}$Rest-frame equivalent width of the O~VI $\lambda$1032 transition, the 
stronger line of the doublet. For {\it FUSE} observations, $W_{\rm r}$ is the 
mean of the measurements from the individual {\it FUSE} channels, each weighted 
inversely by its variance. Similarly, the weighted mean is reported for cases 
where the line is recorded independently in STIS and {\it FUSE} spectra. \\
$^{c}$O~VI absorption lines which show clear evidence of component structure 
(such as the system shown in Figure 2) and which were consequently fitted with 
multiple components are referred to as multicomponent absorbers. O~VI profiles 
adequated fitted with just one line are referred to as single-component systems. 
Parentheses indicate that the system is considered a ``possible'' O~VI detection 
because only one line of the O~VI doublet is detected; the other line is either 
lost in a blend or too weak to detect. \\
$^{d}$Reference: 1. Tripp et al. (2002), 2. Chen \& Prochaska (2000), 3. Savage 
et al. (2001), 4. Tripp \& Savage (2000), 5. Sembach et al. (2001), 6. Tripp et 
al. (2001), 7. Oegerle et al. (2000), 8. Tripp, Savage, \& Jenkins (2000). \\
$^{e}$A Lyman limit absorber showing many metal absorption lines is present at 
this redshift. However,the O~VI $\lambda$1032 line is blended with a strongly 
saturated Milky Way Si~II transition and cannot be measured. A well-detected 
line is present at the expected wavelength of the O~VI $\lambda 1038$ 
transition; this feature has $W_{\rm r} = 96\pm 10$ m\AA . \\
$^{f}$This O~VI doublet only shows one component (see Figure 1). However, it is 
only 340 km s$^{-1}$ from the $z_{\rm abs}$ = 0.22497 absorber, and it is 
possible that all of these lines are all part of the same system. Consequently, 
this case is considered part of a multicomponent system.
\end{table}

\begin{figure}[h]
\begin{center}
\epsfxsize=5.5in
\leavevmode
\epsffile{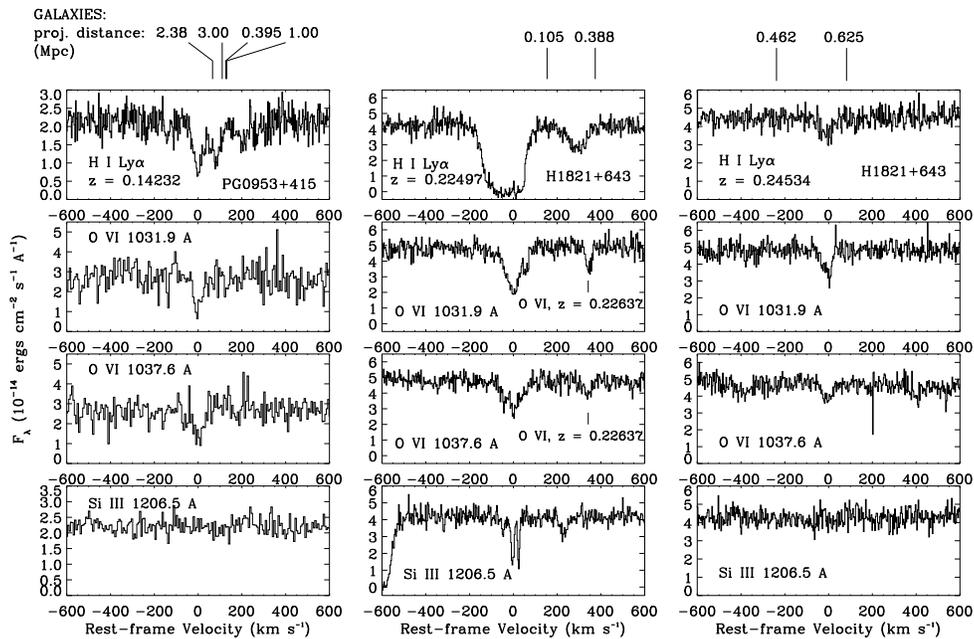}
\end{center}
\caption{\small Examples of intervening O~VI absorbers detected in the STIS 
echelle spectra of PG0953+415 (left panels; see Tripp \& Savage 2000) and 
H1821+643 (center and right panels; Tripp, Savage, \& Jenkins 2000) plotted vs. 
rest-frame velocity ($v = 0$ at the $z_{\rm abs}$ indicated in each panel). The 
velocities of galaxies near these absorbers and their projected distances from 
the sight line are indicated at the top of each panel.}
\end{figure}

\begin{figure}[h]
\begin{center}
\epsfxsize=5.0in
\leavevmode
\epsffile{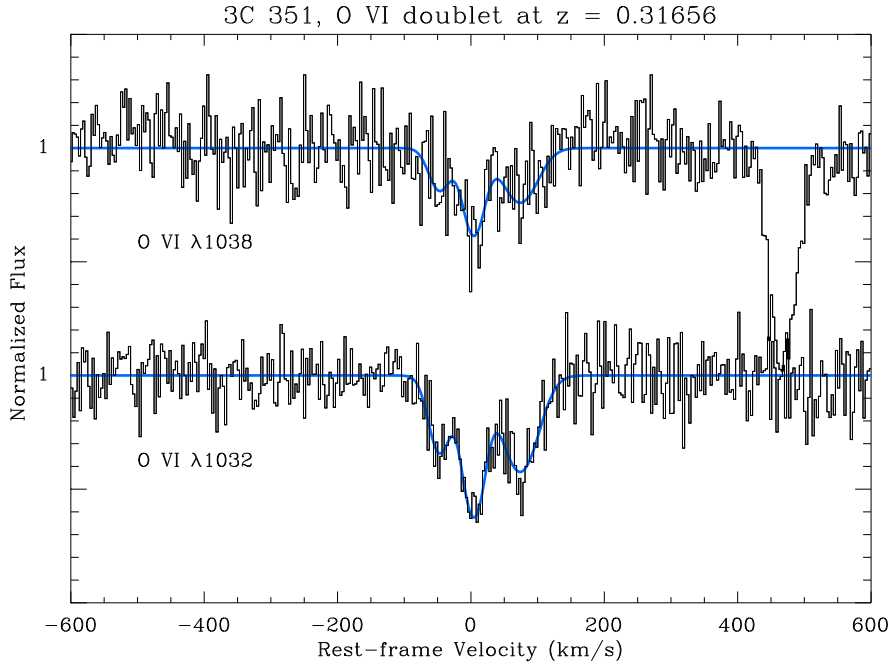}
\end{center}
\caption{\small Example of an O~VI system from the STIS echelle spectrum of 
3C351 (Tripp et al. 2002), plotted vs. rest-frame velocity.}
\end{figure}

The observations and data reduction procedures have been summarized in several 
papers (see references in Table 1). In brief, observations of the QSOs have been 
obtained with STIS (Kimble et al. 1998; Woodgate et al. 1998) and {\it FUSE} 
(Moos et al. 2000; Sahnow et al. 2000). The primary STIS observations employed the E140M 
echelle mode, which provides a resolution of $R = \lambda /\Delta \lambda = 
46,000$ (7 km s$^{-1}$ FWHM) and wavelength coverage from $\sim$1150 to 1700 \AA 
. Some first-order grating observations with STIS have also been obtained, 
primarily to search for corresponding C~IV at longer wavelengths. {\it FUSE} has 
four co-aligned telescopes and Rowland spectrographs which record spectra on two 
microchannel plate detectors. Different coatings in the various {\it FUSE} 
channels enable spectral coverage from 905$-$1187 \AA . The {\it FUSE} 
resolution is estimated to range from 17$-$25 km s$^{-1}$ (FWHM) depending on 
the wavelength and channel. Note that there is some overlap in the wavelength 
coverage of STIS and {\it FUSE}, and in a couple of cases O~VI lines have been 
independently detected by both instruments.

We have also measured galaxy redshifts in the fields of several of the target 
QSOs with Hydra, the fiber-fed multiobject spectrograph on the WIYN telescope, 
in order to learn about the relationships between the low$-z$ absorbers and 
galaxies/environment. The observational and measurement techniques are described 
in Tripp, Lu, \& Savage (1998). A single Hydra setup covers a one-degree field 
of view; we usually surveyed a somewhat larger field by offsetting the field 
center for different fiber configurations.

\section{O~VI Absorber Properties}

\begin{figure}[h]
\begin{center}
\epsfxsize=5.0in
\leavevmode
\epsffile{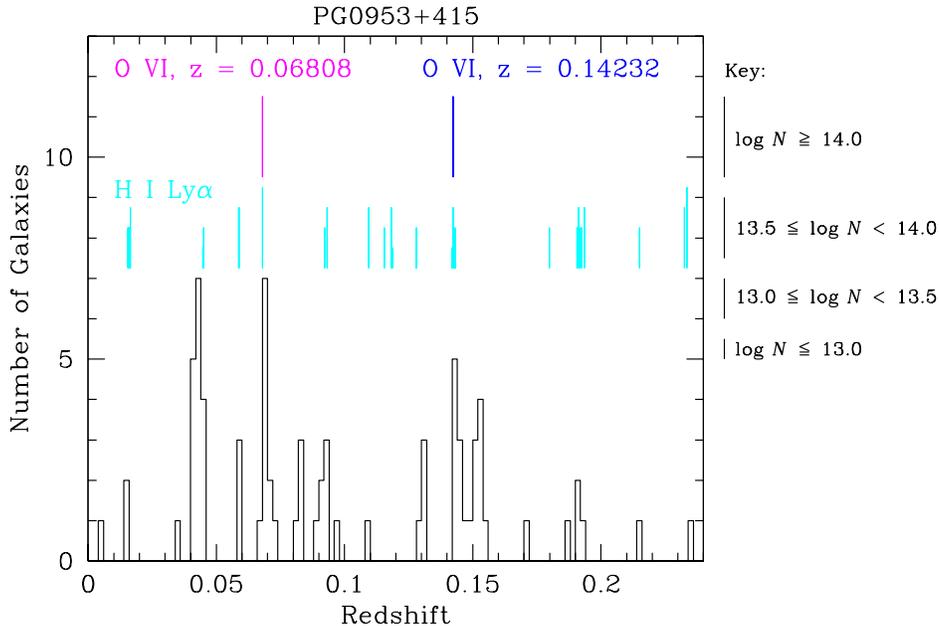}
\end{center}
\caption{\small Redshift distribution of galaxies in the $\sim 1^{\circ}$ field 
centered on PG0953+415. The redshifts of the O~VI absorption systems discovered 
by Tripp \& Savage (2000) and Savage et al. (2001) are indicated at the top of 
the figure, and the Ly$\alpha$ line redshifts are shown below the O~VI 
redshifts. The height of the lines reflects the column density as indicated in 
the key.\label{galreds}}
\end{figure}

Some examples of O~VI absorption lines (and detections, or lack thereof, of 
other species at the same redshift) are shown in Figures 1 and 2. Initial 
results on these absorption systems include the following:

{\it The number of O~VI absorbers per unit redshift ($dN/dz$) is remarkably 
high, and it is likely that these absorption systems trace an important baryon 
reservoir.} However, the contribution to $\Omega _{\rm b}$ is still highly 
uncertain, and there are several outstanding issues, such as whether there is 
substantial double-counting in the baryon census due to overlap between O~VI 
systems and plain Ly$\alpha$ absorbers (see further discussion in \S 4).

{\it The O~VI absorbers are {\bf intervening}.} Excluding systems within 5000 km 
s$^{-1}$ of the QSO redshift, the O~VI systems are located close to luminous 
galaxies. Toward H1821+643, the five definite intervening O~VI systems (Tripp et 
al. 2000, 2001) are within a projected distance of 1 $h_{75}^{-1}$ Mpc or less 
of at least one galaxy with $\mid \Delta v\mid \ = \mid c(z_{\rm gal} - z_{\rm 
abs})/(1 + z_{\rm mean}) \mid \ \leq$ 300 km s$^{-1}$. In some cases, multiple 
galaxies are close to the sight line (see, e.g., the top of Figure 1 or Figure 3 
in Tripp et al. 2001). Combined with the large redshift difference from the QSO, 
this indicates that these are probably intervening systems that trace the 
large-scale gaseous environment in galaxy structures and the IGM rather than 
{\it intrinsic} absorbers which are sometimes ejected to remarkably large 
$\Delta v$ (Hamann \& Ferland 1999, and references therein). 

Similarly, Figure 3 shows the locations in redshift space of the O~VI absorbers 
reported by Tripp \& Savage (2000) and Savage et al. (2001) in the spectrum of 
PG0953+415 with respect to the observed galaxy distribution. Clearly, both the 
O~VI and Ly$\alpha$ absorption lines follow the galaxy distribution. However, 
there are no galaxies particularly close to the O~VI system at $z_{\rm abs}$ = 
0.06807; the closest galaxy is at projected distance of 917 $h_{75}^{-1}$ kpc. 
This may be due to incompleteness of the galaxy redshift survey, but this is 
also predicted by the cosmological simulations, which find that the O~VI 
absorbers arise in low-overdensity, unvirialized regions, i.e., the lines are 
expected to be found in large-scale galaxy filaments, but there may not be a 
galaxy very nearby (see, e.g., Cen et al. 2001).

{\it The physical conditions and metallicities of the O~VI absorbers are highly 
variable, and the conditions are frequently complex.} In the intervening systems 
identified so far, the H~I Ly$\alpha$ transition is always detected along with 
the O~VI lines, as expected for gas with sub-solar metallicity which is 
collisionally ionized at $T < 10^{6}$ K or which is photoionized. In some cases 
these are the only detected absorption lines, but in other cases lines due to 
low and intermediate ionization stages (e.g., Si~II, Si~III, N~V, C~III, C~IV) 
are also detected, usually in systems with higher H~I column densities. The 
O~VI/H~I column density ratio varies substantially from system to system (e.g., Tripp 
et al. 2000); this may be due to significantly different metallicities, physical 
conditions, or both. Many systems show complicated component structure (see, 
e.g., Figure 1, Figure 2 in Tripp et al. 2000, or Figure 1 in Tripp et al. 
2001), and there is often strong evidence that the absorbing media are 
multiphase. For example, the O~VI and Si~III absorption lines at $z_{\rm abs}$ = 
0.225 toward H1821+643 (middle panels in Figure 1) cannot be explained by a 
single-phase absorber.

\section{Baryonic Content}

\begin{figure}[h]
\begin{center}
\epsfxsize=5.0in
\leavevmode
\epsffile{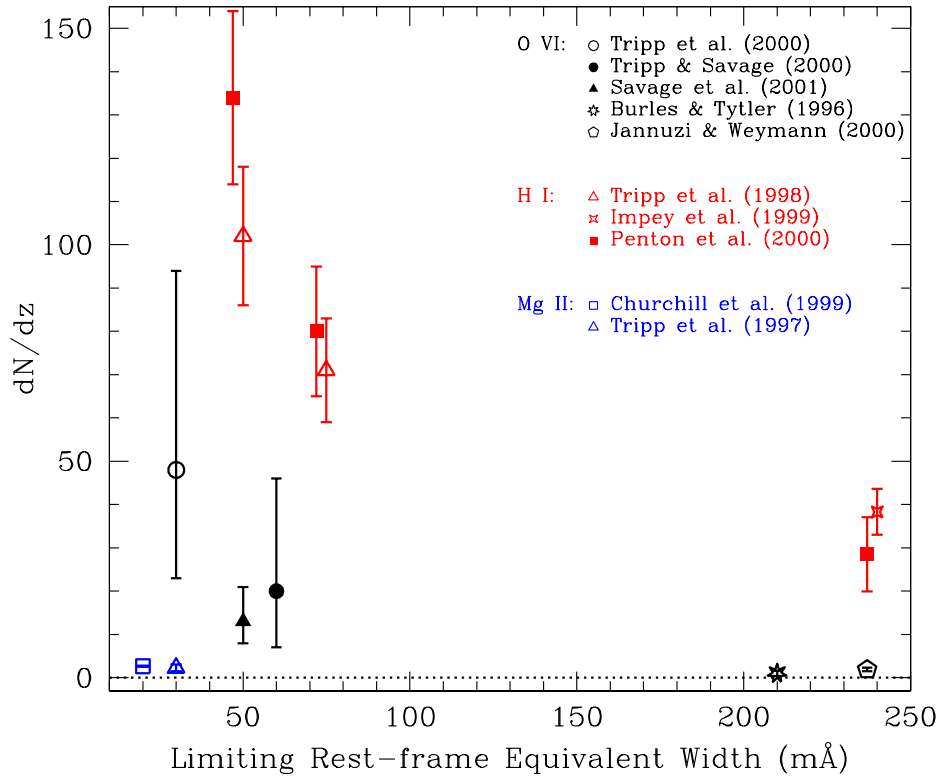}
\end{center}
\caption{\small Number of absorbers per unit redshift ($dN/dz$) vs. limiting 
equivalent width of the sample for O~VI, H~I, and Mg~II systems (as indicated in 
the key). Note that the Burles \& Tytler and Jannuzi \& Weymann measurements as 
well as the Mg~II points were derived from somewhat higher redshift data, and 
$dN/dz$ is expected to depend on redshift. However, Tripp et al. (2002) find the 
$dN/dz$ of O~VI systems decreases with increasing limiting equivalent width at 
the lowest redshifts as well.}
\end{figure}

It has been noted for some time that there is a ``missing baryon problem'' at 
the present epoch: the well-observed baryon repositories in the nearby universe 
(e.g. Fukugita et al. 1998) fail to provide the quantity of baryons expected 
based on D/H measurements (e.g., Burles \& Tytler 1998) or observations of the 
high$-z$ Ly$\alpha$ forest (e.g., Rauch et al. 1997; Weinberg et al. 1997; 
Schaye 2001). As noted in \S 1, cosmological simulations suggest that the 
missing baryons are hidden in low-density regions of the IGM that have been 
shock-heated to $10^{5} - 10^{7}$ K. Such gas can be revealed using QSO 
absorption lines such as the O~VI or Ne~VIII doublets (the O~VI ion fraction 
peaks at $T \sim$ 300,000 K), and this has been the primary motivation for the 
observations discussed in this paper.

The first STIS observations for this program (Tripp et al. 2000; Tripp \& Savage 
2000) suggested that the number of O~VI systems per unit redshift is indeed 
interesting. Tripp et al. (2000) found $dN/dz \sim$ 50 for $W_{\rm r} >$ 30 m\AA 
, albeit with large error bars. This is more comparable to $dN/dz$ of low$-z$ 
Ly$\alpha$ absorbers than of other types of metal absorbers, as shown in Figure 
4. We now have more observations, and Figure 4 summarizes the $dN/dz$ of O~VI 
absorbers from more recent papers, as a function of the limiting equivalent 
width of the particular sample. The redshift density of other types of absorbers 
from the literature are also shown in Figure 4 for comparison. Figure 4 shows 
that $dN/dz$(O~VI) appears to increase with decreasing limiting equivalent 
width,\footnote{Note that the Burles \& Tytler (1996), Jannuzi \& Weymann 
(2000), and the Mg~II results apply to somewhat higher redshifts, and therefore 
this figure does not provide a completely fair comparison. However, Tripp et al. 
(2002) find that $dN/dz$(O~VI) decreases with increasing limiting $W_{\rm r}$ at 
lower redshifts as well.} as predicted by cosmological simultations (Cen et al. 
2001; Fang \& Bryan 2001). However, more observations are needed to firmly 
establish this trend.

The high $dN/dz$ suggests that O~VI systems are an important baryon reservoir. 
Following analogous calculations (e.g., Burles \& Tytler 1996; Rao \& Turnshek 
2000), the baryonic content of the O~VI absorbers, expressed in the usual 
fashion ($\Omega = \rho /\rho _{\rm c}$), can be estimated using the following 
expressions:
\begin{equation}
\Omega _{\rm b}({\rm O \ VI}) = \frac{\mu m_{\rm H} H_{0}}{\rho _{c} c 
f({\rm O \ VI})} \left( \frac{\rm O}{\rm H} \right)^{-1}_{\rm O \ VI} 
\frac{\sum_{i} N_{i}({\rm O \ VI})}{\sum_{i} \Delta X_{i}}
\end{equation}
or, alternatively,
\begin{equation}
\Omega _{\rm b}({\rm O \ VI}) = \frac{\mu m_{\rm H} H_{0}}{\rho _{c} c 
f({\rm O \ VI})} \left( \frac{\rm O}{\rm H} \right)^{-1}_{\rm O \ VI} 
\left(\frac{dN}{dz} \right) \frac{<N_{\rm O~VI}>}{(1+z)}
\end{equation}
assuming $q_{0}$ = 0. Here $\Delta X_{i}$ is the absorption distance interval 
(e.g., Burles \& Tytler 1996) probed to the $i^{\rm th}$ sight line (corrected 
for spectral regions blocked by ISM or extragalactic lines), $\mu$ is the mean 
atomic weight, (O/H)$_{\rm O~VI}$ is the mean oxygen abundance by number, 
$<N_{\rm O~VI}>$ is the mean O~VI column density of the sample, and $f$(O~VI) is 
the ion fraction. Values for (O/H)$_{\rm O~VI}$ and $f$(O~VI) must be assumed, 
but a conservative result can be obtained by assuming the largest plausible 
values for these parameters thereby minimizing $\Omega _{\rm b}$(O~VI). In this 
way, Tripp et al. (2000) derived $\Omega_{\rm b}$(O~VI) $\geq 0.004 h_{75}^{-1}$ 
for a sample with $W_{\rm r} \geq 30$ m\AA , assuming a mean metallicity of 1/10 
solar and $f$(O~VI) = 0.2. Similarly, Tripp \& Savage (2000) obtained 
$\Omega_{\rm b}$(O~VI) $\geq 0.003 h_{75}^{-1}$ for a sample with $W_{\rm r} 
\geq 60$ m\AA . These lower limits are comparable to the cosmological mass 
density in the form of stars, cool neutral gas, and X-ray emitting cluster gas 
at the present epoch (Fukugita, Hogan, \& Peebles 1998).

However, there are several issues and sources of uncertainty. First, the results 
have large error bars due to the small size of the initial samples; using eqn. 
2, Tripp \& Savage (2000) obtained $\Omega_{\rm b}$(O~VI) $\geq 
0.003^{+0.004}_{-0.002} h_{75}^{-1}$ at the $1\sigma$ level. Therefore it may 
not be too surprising that more recent estimates of $\Omega_{\rm b}$(O~VI) with 
larger samples yield somewhat smaller values: Savage et al. (2001) find 
$\Omega_{\rm b}$(O~VI) $\geq 0.002 h_{75}^{-1}$ with a limiting equivalent width 
of 50 m\AA , for example. Second, the blocking of some regions of the spectrum 
by unrelated lines (e.g., due to the Milky Way ISM) reduces the redshift path 
probed for O~VI lines. The blocking correction affects $\Omega_{\rm b}$(O~VI) 
because reducing $\Delta z$ (and $\Delta X$ in eqn. 1) increases $dN/dz$ and 
$\Omega_{\rm b}$(O~VI). In the initial papers, we used a very simple algorithm 
to estimate this blocking correction. Any strong line that obscured either the 
O~VI $\lambda$1032 or $\lambda$1038 lines reduced the total $\Delta z$. This is 
conservative and simple, but since we are using a doublet, a more accurate 
blocking correction can be obtained by checking for the other line of the 
doublet. If a particular strong line masks $\lambda$1032 at some redshift, a 
corresponding $\lambda$1038 feature should be present at the same $z$ (and vice 
versa). If no corresponding feature is evident, then the strong line does not 
block an O~VI system. A more accurate assessment of blocking increases $\Delta 
z$ and decreases $\Omega_{\rm b}$(O~VI), but the change is less than a factor of 
2. Third, the mean metallicity of the absorbers may be higher than 0.1${\rm 
Z}_{\odot}$. However, this is countervailed by the fact that $f$(O~VI) is almost 
certainly less than 0.2. From their cosmological simulations, Cen et al. (2001) 
find $f$(O~VI)(Z/${\rm Z}_{\odot}) \sim$ 0.01 for a collisionally ionized O~VI 
absorber with a 50 m\AA\ $\lambda$1032 line. This is a factor of $\sim$2 lower 
than the value we have assumed [reducing $f$(O~VI)(Z/${\rm Z}_{\odot})$ 
increases $\Omega_{\rm b}$(O~VI)]. Fourth, there is a concern about 
double-counting in the baryon census. The contribution of low$-z$ Ly$\alpha$ 
absorbers to the baryon inventory has been estimated (Penton et al. 2000). 
These absorbers are assumed to be photoionized and relatively cool 
($T \sim 10^{4}$ K). However, in the hydrodynamic simultions, some Ly$\alpha$ 
lines arise in substantially hotter gas (see Figure 4 in Dav\'{e} \& Tripp 
2001). Likewise, some O~VI systems could be photoionized by the UV background 
from QSOs if the density is low enough. Consequently, some effort is required 
to disentangle the photoionized and shock-heated absorbers for purposes of the 
baryon census. This will likely be challenging, but guidance is already being 
provided by simulations (Cen et al. 2001; Fang \& Bryan 2001).

\section{Physical Conditions and Metallicity}

\begin{figure}[h]
\begin{center}
\epsfxsize=5.0in
\leavevmode
\epsffile{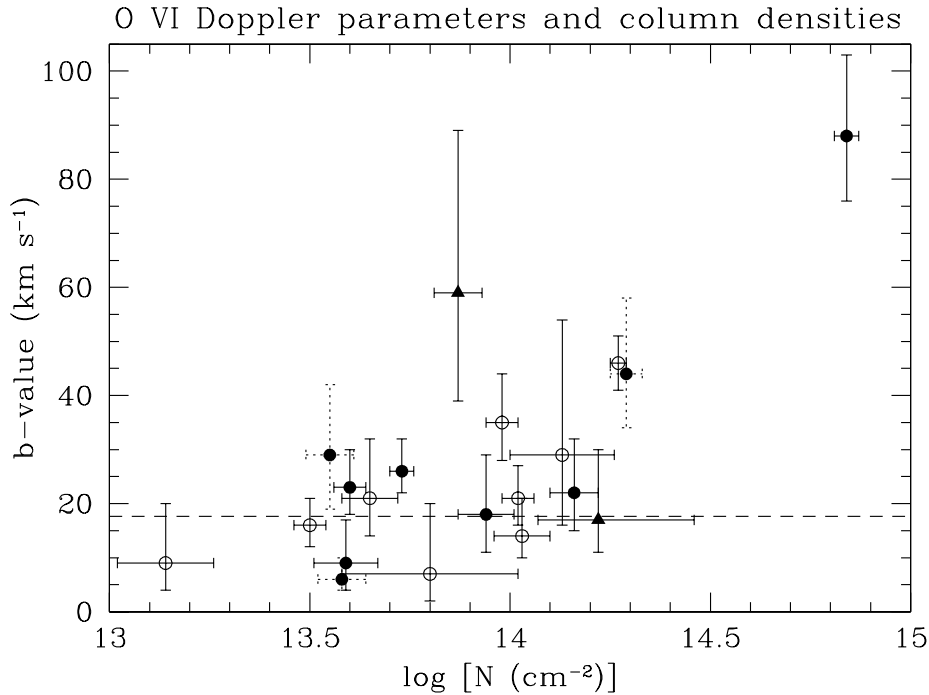}
\end{center}
\caption{\small Column densities and Doppler parameters of intervening O~VI 
absorbers measured with the Voigt profile fitting software of Fitzpatrick \& 
Spitzer (1997). Absorbers fit with a single component are indicated with filled 
symbols, and the column densities and $b-$values of individual components of 
multicomponent absorbers (such as the system shown in Figure 2) are shown with 
open symbols. Measurements from STIS observations are plotted with circles, and 
the triangles are from {\it FUSE} data. Definite O~VI absorbers have solid error 
bars, and ``possible'' O~VI systems (see Table 1) are shown with dotted error 
bars. The horizontal dashed line indicates the O~VI $b-$value at $T = 300,000$K, 
if the line broadening is dominated by thermal motions.}
\end{figure}

As noted in the previous sections, it is important to understand the physical 
conditions and metallicities of the O~VI systems in order to assess their 
baryonic content. What fraction of the O~VI (and Ly$\alpha$) absorbers are 
photoionized? How can the photoionized and collisionally ionized systems be 
distinguished? Answering these questions will greatly alleviate the 
double-counting problem. What metallicity should be assumed for calculation of 
$\Omega _{\rm b}$(O~VI)? Physical conditions and metallicity are interesting 
properties in their own right, for understanding the nature and evolution of 
the IGM and its influence on and interaction with galaxies and large-scale 
structures.

Our group has scrutinized the ionization mechanism in several O~VI 
absorbers and derived metallicity constraints (e.g., Tripp \& Savage 
2000; Tripp et al. 2001; Savage et al. 2001). It is often difficult 
to definitely rule out one source of ionization or another, but in 
several cases one ionization mechanism is strongly favored. For 
example, in the intervening O~VI absorber at $z_{\rm abs}$ = 0.06807 
in the spectrum of PG0953+415, H~I, C~III, C~IV, and N~V lines are 
detected and are well-aligned with the O~VI lines, and all of the 
column densities are reproduced by a photoionization model with 
metallicity ${\rm Z} = 0.4^{+0.6}_{-0.2}{\rm Z_{\odot}}$ (Savage et 
al. 2001). Perhaps more importantly, the H~I Ly$\alpha$ and 
Ly$\beta$ lines associated with the O~VI are narrow and appear to be 
composed of only one component, and the upper limit on $T$ from the 
width of the H~I lines is $T \leq 4.1 \times 10^{4}$ K. This 
precludes collisional ionization, at least in 
equilibrium,\footnote{It is alternatively possible, if the density 
is high enough, that the gas was initially hot enough to be 
collisionally ionized but then cooled more rapidly than it 
recombined leading to overionized, cool gas (see, e.g., Edgar \& Chevalier 
1986).} unless the H~I 
arises in a different phase from the O~VI. A similar situation is 
found for the O~VI system studied by Tripp \& Savage (2000), 
although in this case it is possible to hide a broad component in 
the complicated Ly$\alpha$ profile, assuming the absorber is a 
multiphase medium (see their Figure 6). The O~VI/Ly$\alpha$ system 
at $z_{\rm abs} = 0.1212$ in the spectrum of H1821+643 is an example 
of a system which is likely collisionally ionized (Tripp et al. 
2001). In this case a very broad component is evident in the 
Ly$\alpha$ profile, which implies that the gas has $T \leq 10^{5.6}$ 
K. In addition, C~IV is {\bf not} significantly detected, and the O~VI/C~IV 
column density ratio provides a lower limit on $T$. Assuming the absorber 
is collisionally ionized and in equilibrium, Tripp et al. derive 
$10^{5.3} \leq T \leq 10^{5.6}$ K and $-1.8 \leq$ [O/H] $\leq -
0.6$.\footnote{[O/H] = log (O/H) - log (O/H)$_{\odot}$.}

The $b-$values of the lines provide an important clue regarding the physical 
conditions of the absorbers. They provide a straightforward upper limit on the 
temperature of the gas, $T \leq mb^{2}/2k$ (an upper limit because other factors 
besides thermal motions, such as turbulence or unresolved blended components, 
can contribute to the line width). The Ly$\alpha$ lines provide the most useful 
constraint along these lines due to the lower mass of hydrogen. However, Fang \& 
Bryan (2001) and Cen et al. (2001) have recently noted that in their 
cosmological simulations, the photoionized and collisionally ionized O~VI 
systems can be distinguished on the basis of the O~VI $b-$values: the 
photoionized absorbers are narrower and have lower equivalent widths than the 
collisionally ionized systems. Figure 5 shows the Doppler parameters and column 
densities of the O~VI absorbers in Table~1, determined using the profile-fitting 
software written by Fitzpatrick \& Spitzer (1997) with the line-spread functions 
from the STIS Handbook, along with the formal 1$\sigma$ error bars from the 
profile-fitting code. The Doppler parameter of an O~VI line arising in gas at $T 
= 300,000$ K, assuming the line width is dominated by thermal broadening, is 
shown with a dashed line. The median $b-$value of the entire sample is 22 km 
s$^{-1}$, and the median O~VI column density is $8.7 \times 10^{13}$ cm$^{-2}$. 
Of course, some of the ``single-component'' systems are so broad that we can be 
confident that the broadening is not entirely due to thermal motions (e.g., the 
highest column density system in the figure), but we do not have sufficient 
information in the line profiles to decompose these cases into subcomponents. 
Nevertheless, it is intriguing that the majority of the lines shown in Figure~5 
have $b-$values reasonably close to the value expected for gas at $T \sim 
300,000$ K, the temperature at which O~VI peaks in abundance. There are a few 
apparently narrow O~VI lines (but usually these are consistent with $T \sim 
300,000$ K within the 1$\sigma$ error bars), and these tend to be weaker 
systems. This suggests that most of the O~VI absorbers arise in collisionally 
ionized hot gas.

\section{Summary}

This contribution has briefly summarized results obtained from high-resolution 
spectroscopy of low-redshift O~VI absorbers. Initial observations indicate that 
there is a high number of O~VI systems per unit redshift in the nearby universe, 
and these systems likely harbor a significant quantity of baryons. More 
observations are planned to increase the sample size and total redshift path and 
thereby place these measurements on firmer statistical grounds. The systems are 
correlated with luminous galaxies and are intervening, not ejected/intrinsic 
absorbers. However, the nature of the relationship between galaxies and O~VI 
systems is not yet clear -- these lines could arise in unvirialized large-scale 
filaments (Cen et al. 2001), the intragroup medium in galaxy groups (Mulchaey et 
al. 1996), escaping winds from individual galaxies (Heckman et al. 2001), or the 
bound ISM of individual galaxies (Jenkins 1978; Savage et al. 2000). The 
physical conditions and metallicities show a broad range of properties, but in 
some cases the absorbing media are multiphase. In general, the properties of the 
O~VI absorbers appear to be broadly consistent with predictions of cosmological 
simulations of structure growth, but much more observational work and 
improvements in the simulations are required to flesh out the nature of these 
absorption line systems and accurately assess their contribution to the baryon 
budget.

\acknowledgements

I thank John Mulchaey and John Stocke for organizing a stimulating meeting in 
tribute to Ray Weymann, in which it was an honor to participate. I also thank my 
numerous collaborators for their hard work and crucial contributions. This 
research was supported by NASA through grants GO-08165.01-97A and GO-08695.01-A 
from the Space Telescope Science Institute as well as NASA Grant NAS5-30110.


\begin{references}
\reference{Burles, S., \& Tytler, D. 1996, \apj, 460, 584}
\reference{Burles, S., \& Tytler, D. 1998, \apj, 499, 699}
\reference{Cen, R., \& Ostriker, J. P. 1999, \apj, 514, 1}
\reference{Cen, R., Tripp, T. M., Ostriker, J. P., \& Jenkins, E. B. 2001, 
\apjl, submitted (astro-ph/0106204)}
\reference{Chen, H.-W., \& Prochaska, J. X. 2000, \apj, 543, L9}
\reference{Churchill, C. W., Rigby, J. R., Charlton, J. C., \& Vogt, S. S. 1999, 
\apjs, 120, 51}
\reference{Dav\'{e}. R., Cen, R., Ostriker, J. P., et al. 2001, \apj, 552, 473}
\reference{Dav\'{e}, R. \& Tripp, T. M. 2001, \apj, 553, 528}
\reference{Edgar, R., \& Chevalier, R. 1986, \apj, 310, L27}
\reference{Fang, T., \& Bryan, G. L. 2001, \apjl, submitted}
\reference{Fitzpatrick, E. L., \& Spitzer, L. 1997, \apj, 475, 623}
\reference{Fukugita, M., Hogan, C. J., \& Peebles, P. J. E. 1998, \apj, 503, 
518}
\reference{Hamann, F., \& Ferland, G. 1999 \araa, 37, 487}
\reference{Heckman, T. M., Sembach, K. R., Meurer, G. R., Strickland, D. K., 
Martin, C. L., Calzetti, D., \& Leitherer, C. 2001, \apj, 554, 1021}
\reference{Impey, C. D., Petry, C. E., \& Flint, K. P. 1999, \apj, 524, 536}
\reference{Jannuzi, B. T., \& Weymann, R. J. 2000, private communication}
\reference{Jenkins, E. B. 1978, \apj, 219, 845}
\reference{Kimble, R. A., Woodgate, B. E., Bowers, C. W., et al. 1998, \apj, 
492, L83}
\reference{Moos, H. W., Cash, W. C., Cowie, L. L., et al. 2000, \apj, 538, L1}
\reference{Mulchaey, J. S., Mushotzky, R. F., Burstein, D., \& Davis, D. S. 
1996, \apj, 456, L5}
\reference{Oegerle, W. R., Tripp, T. M., Sembach, K. R., et al. 2000, \apj, 538, 
L23}
\reference{Penton, S. V., Shull, J. M., \& Stocke, J. T. 2000, \apj, 544, 150}
\reference{Rao, S. M., \& Turnshek, D. A. 2000, \apjs, 130, 1}
\reference{Rauch, M., Miralda-Escud\'{e}, J., Sargent, W. L. W., et al. 1997, 
\apj, 489, 7}
\reference{Sahnow, D. J., Moos, H. W., Ake, T. B., et al. 2000, \apj, 538, L7}
\reference{Savage, B. D., Sembach, K. R., Jenkins, E. B., et al. 2000, \apj, 
538, L27}
\reference{Savage, B. D., Sembach, K. R., Tripp, T. M., \& Richter, P. 2001, 
\apj, submitted}
\reference{Schaye, J. 2001, \apj, in press (astro-ph/0104272)}
\reference{Sembach, K. R., Howk, J. C., Savage, B. D., Shull, J. M., \& Oegerle, 
W. R. 2001, \apj, in press}
\reference{Tripp, T. M., Giroux, M. L., Stocke, J. T., Tumlinson, J., \& 
Oegerle, W. R. 2001, \apj, in press}
\reference{Tripp, T. M., Lu, L., \& Savage, B. D. 1997, \apjs, 112, 1}
\reference{Tripp, T. M., Lu, L., \& Savage, B. D. 1998, \apj, 508, 200}
\reference{Tripp, T. M., \& Savage, B. D. 2000, \apj, 542, 42}
\reference{Tripp, T. M., Savage, B. D., \& Jenkins, E. B. 2000, 
\apj, 534, L1}
\reference{Tripp, T. M., et al. 2002, in preparation}
\reference{Weinberg, D. H., Hernquist, L., Miralda-Escud\'{e}, J., \& Katz, N. 
1997, \apj, 490, 564}
\reference{Woodgate, B. E., Kimble, R. A., Bowers, C. W., et al. 1998, \pasp, 
110, 1183}
\end{references}
\end{document}